# On the Carbon Dioxide Capture by Quaternary Ammonium-Based and Phosphonium-Based Ionic Liquids. The Role of Steric Hindrances and Transition States

Vitaly V. Chaban[1]

(1) P.E.S., Vasilievsky Island, Saint Petersburg 190000, Russian Federation.

**Abstract**. Global warming is seen as a drastic environmental problem nowadays. Carbon dioxide ($CO_2$) concentration in the Earth's atmosphere is linked to the average temperature on the surface of the planet. Carbon capture and storage is an important technological endeavor aiming to improve the ecology. The present work investigates reaction paths that are responsible for $CO_2$ chemisorption by the ammonium- and phosphonium-based ionic liquids containing an aprotic heterocyclic anion 2-cyanopyrrolidine. We show that two moles of $CO_2$ per one mole of the gas scavenger can be theoretically fixed by such ionic liquids. Both the cation and anion participate in the chemisorption. The corresponding standard enthalpies are moderately negative. The barriers of all reactions involving the phosphonium-based cation are relatively small and favor practical applications of the considered sorbents. The performance of the ammonium-based cation is less favorable due to the inherent instability of the tetraalkylammonium ylide. The role is phosphonium ylide in the mechanism of the reaction is carefully characterized. The reported results foster a fundamental understanding of the outstanding $CO_2$ sorption performance of the quaternary ammonium and phosphonium-based 2-cyanopyrrolidines.

**Graphical Abstract**

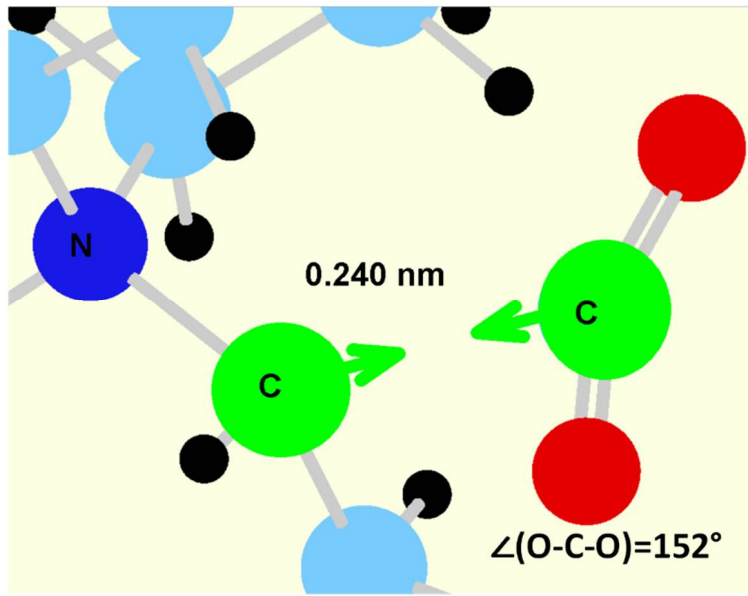

## Introduction

The term of global warming stands for the designation of the large-scale trend.[1] During the last twelve decades, the average Earth's temperature increased by 0.08 degrees Celsius per year. Furthermore, during the last forty years, the pace of this change increased more than twice. Although the recorded changes may seem insignificant for a non-specialist, they do have ecological consequences including thawing of glaciers, climate changes, more acidic oceans, etc.[2-4] Human technologies lead to higher emissions of carbon dioxide ($CO_2$) that may be the cause of environmental problems. The technological superstructures are currently being developed aiming to capture, store and chemically modify $CO_2$ to obtain useful and/or benign products.[5,6]

$CO_2$ is a highly thermodynamically stable compound that is a final product of organic matter combustion. Its stability is a major reason why $CO_2$ scavenging is a challenging chemical endeavor that will likely never get a straightforward solution. Nevertheless, novel robust methods for $CO_2$ processing[7-15] can be elaborated and industrially implemented.[5] Thoughtful application and fine-tuning of the versatile nanoscale pores play a paramount role in the gradual increase of the presently available $CO_2$ sorption capacities.[16-19] Apart from the up-to-date breakthrough developments, amine scrubbing is a leading and well-established technology nowadays, even though it presents evident drawbacks, such as perpetual degeneration of the amine solutions due to their working chemical reactions.[20,21] Time is necessary for humanity to implement more chemically advanced solutions on the large scale.

Room-temperature ionic liquids (RTILs) represent a huge-sized class of organic-inorganic chemical compounds that are liquid at standard conditions.[22-28] The liquid state of the prospective $CO_2$ sorbent is a definite advantage since it allows the gas molecules to spontaneously diffuse to the location where chemisorption takes place. Furthermore, RTILs exhibit negligible volatility even in their pure state thanks to a strong cation-anion attraction.

An ideal $CO_2$ sorbent would allow reversible chemisorption so that the fixed gas could be extracted and the working liquid could be regenerated. RTILs contain chemically tunable structures meaning that new compounds can be synthesized on request in a relatively easy way. Fine-tunable physical-chemical properties form a prerequisite to developing task-specific systems and environments. Computer simulations of RTILs constitute a large, vivid, and vigorously developing research field with substantial methodological and engineering advances.[29-33] Padua and coworkers[32] recently developed a highly efficient polarizable force field to simulate condensed phases of ionic liquids. In turn, Canongia Lopes and coworkers made a considerable impact on the understanding of versatile aqueous ionic liquid-based systems through the simulation of peculiar non-covalent interactions.[30,31] Cordeiro and coworkers contributed an advanced description of the ionic liquids' behavior at the interface and linked new insights to the differential capacitance of a possible electrochemical device.[33] Andreeva and coworkers investigated thermodynamics for a large number of ionic liquid families and outlined their suitability to act as non-volatile and non-flammable $CO_2$ sorbents.[34,35]

The most promising RTILs in the context of $CO_2$ sorption are composed of bulky organic cations and chemically reactive anions.[23,25,27] The aprotic heterocyclic anions (AHAs) represent an interesting family of organic structures with a set of potential applications in different fields of chemical engineering due to their strong affinity to the proton. This feature of AHAs can arguably convert many thermodynamically forbidden reactions into favorable ones. It is essential to couple AHAs with the bulky and asymmetric cations to block all strong electrostatic attractions in the resulting liquid-state system. The organic structures with the delocalized excess or deficient electronic charge sterically prevent RTILs from crystallizing at the ambient temperature. Higher conformation flexibility of the particles favors their smarter performance gas greenhouse gas scavengers. The inability of the cation and the anion to attain a distinct electrostatically-driven coordination pattern liberate their potential to bind $CO_2$. Upon the development of competitive AHA-based RTILs, there is always an interplay of melting point and shear viscosity that limits out possibilities to tune physical-chemical properties of the prospective $CO_2$ scavengers.[36]

In the present theoretical work, we investigate a few of the most promising ionic liquids in the context of $CO_2$ capture. Indeed, in the tetrabutylammonium 2-cyanopyrrolidine and tetrabutylphosphonium 2-cyanopyrrolidine, both the cation and the anion must reversibly react with $CO_2$. The chemisorption reaction at the anion's reaction site follows the well-known carbamate formation mechanism. In turn, the reaction at the quaternary cation takes place at its α-carbon atom. It is facilitated by the cation's deprotonation leading to the ylide formation (Figure 1).[25] While tetrabutylphosphonium ylide is a stable chemical compound with negative formation energy, tetrabutylammonium ylide is metastable due to the inability of the nitrogen atom to support the valence of five. Nonetheless, the ultimate products are stable in all cases of the chemisorption reactions.

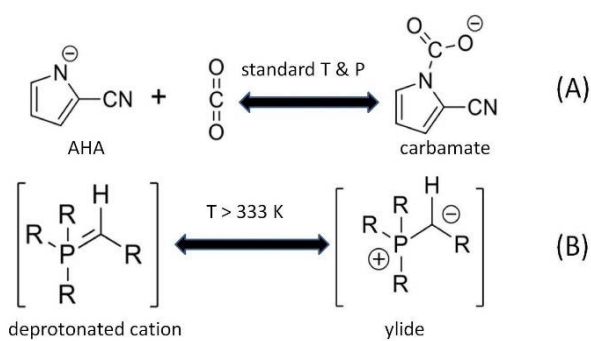

Figure 1. The proposed mechanisms of the two $CO_2$ chemisorption reactions. (A) The formation of carbamate out of the AHA anion. (B) The formation of the ylide molecule out of the quaternary cation after its deprotonation.

The previous experimental works[23,25,27] revealed and confirmed an outstanding $CO_2$ sorption potential of the aprotic heterocyclic anions paired with quaternary cations. In particular, the Brennecke group made a significant impact on this field by introducing a separate class of liquid-state compounds. The RTILs based on the heterocyclic anions exhibit high promise in the $CO_2$ capturing endeavor. They react stoichiometrically, whereas the chemisorption is reversible. Brennecke and coworkers found a solution to the high-viscosity problem of the AHA-based RTILs by adding a certain proportion of tetraglyme.[37] The latter is not viscous, but at the same time, it is relatively non-volatile representing a rare set of physical-chemical properties. In combination with the AHA-containing RTILs, tetraglyme engenders solutions with potentially important and quite unusual physical-chemical properties. The phosphonium-based RTILs paired with aprotic anions form both a carbamate product out of the anion and a carboxyl product out of the cation. Therefore, two moles of $CO_2$ per one mole of the RTIL can be bound.[25]

The phosphonium ylide is an important intermediate in the $CO_2$ chemisorption by the tetraalkylphosphonium-based RTILs. The ylide is a neutral compound that forms reversibly at above 333 K in the case of the chemisorption reaction of tetraalkylphosphonium-based RTILs with carbon dioxide.[25] The presence of $CO_2$ in the system favors the formation of ylide. Compared to the current industrial-scale technology that employs aqueous monoethanolamine solutions, the finely tuned RTILs can substantially decrease the energy demands. According to the very recent quantification performed by Gohndrone and coworkers,[25] the anion's chemisorption is kinetically more favorable, i.e. the major product of $CO_2$ chemisorption by the bulky phosphonium-based ionic liquids is a carbamate.

We presently divide the chemisorption reactions that presumably occur in the systems of "liquid tetrabutylammonium/tetrabutylphosphonium 2-cyanopyrrolidine + gaseous carbon dioxide" type into several sub-processes. Each sub-process includes an independent chemical reaction, i.e. formation of the covalent bonds. We scan the reaction paths of every stage, identify potential barriers, characterize stationary points, and compute thermodynamic properties of the chemisorption-related physical and chemical transformations. The existence of the first-order saddle points along the reaction pathway is confirmed by the presence of a single imaginary frequency in the system's vibrational profile and characterized in terms of geometric parameters and electron density distributions. For the computationally efficient investigation, we select a less bulky cation, as compared to the one used experimentally. However, the investigated length of the alkyl chain is large enough to detect possible sterical hindrances that impact $CO_2$ chemisorption in real-world greenhouse gas capture.

**Methods and Procedures**

The chemical compositions for which the reaction paths were investigated include one cation (tetrabutylphosphonium or tetrabutylammonium), one AHA (2-cyanopyrrolidine), and one $CO_2$ molecule (Figure 2). The simulation of the ionic compounds involves a strong cation-anion attraction that needs to be reproduced for accurate results. For this reason, our systems explicitly include all three essential components participating in the investigated reactions. The maximum size of the systems is limited by the affordable number of the internal variables to avoid the numerical problem upon the analysis of the sampled process.

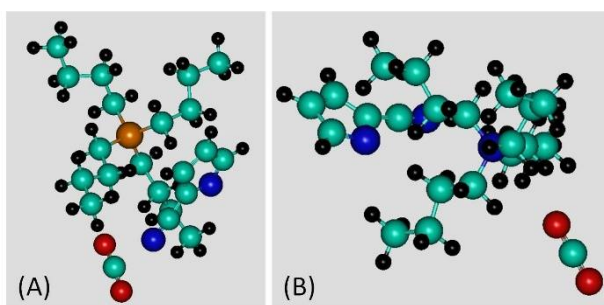

Figure 2. The optimized geometries of the simulated systems. (A) tetrabutylphosphonium 2-cyanopyrrolidine and (B) tetrabutylammonium 2-cyanopyrrolidine. The eigenfollowing procedure of minimizing forces acting on every atom was applied before

the propagation of the reaction paths. The oxygen atoms are red; the nitrogen atoms are blue; the carbon atoms are cyan; the hydrogen atoms are black.

Each chemical composition was employed for the simulation of three reaction paths: (1) proton exchange; (2) $CO_2$ reaction with the α-carbon of the quaternary cation (carboxylation); (3) $CO_2$ reaction with 2-cyanopyrrolidine (carbamate formation). The geometries of all systems were optimized before the stages of the reactions were simulated. The proposed reaction paths were simulated by forcibly moving one of the internal system's coordinates towards the desired value. The selected reaction coordinate was adjusted with a spatial step of 0.001 nm which corresponded to the gradual approach of the prospective reactants to one another. The partial geometry optimization of the system using the eigenfollowing algorithm was performed at every reaction step. The very small step of the reaction coordinate alteration was employed knowingly to record all potential barriers on the way of the chemisorption. Particularly, the sterical potential barriers due to the presence of four alkyl chains were identified. In total, over 1 000 000 self-consistent field calculations (optimizations of the system's wave function) were carried out. The bonded and non-bonded interactions in the systems were described according to the PM7 Hamiltonian.[38] This Hamiltonian includes empirical corrections for all applicable chemical phenomena, such as interatomic dispersion forces, hydrogen bonding, and correlation of the valence electrons.

The self-consistent field convergence criterion upon the iterative optimization of the wave function was set to $0.5 \times 10^{-5}$ kJ mol$^{-1}$ for the two successive cycles of computation. The force minimization convergence criterion was set to 0.5 kJ mol$^{-1}$ nm$^{-1}$ for the gradient norm.

The computed reaction paths provide the smooth total potential energy change upon the internal coordinate change. The ion-molecular configurations corresponding to the potential barriers may be transition states. To ensure that they indeed are such, we re-optimized the corresponding intermediate structures. The vibrational profiles without Raman frequencies were computed to identify imaginary frequencies (one per system) that are direct proofs of the successful saddle point location. The empirical dispersion correction proposed by Grimme and coworkers[39] for ninety-four chemical elements was applied.

The wave function calculations, geometry optimization procedures, and reaction coordinate propagation were performed in a tightly bound set of the computational chemistry programs, in which MOPAC-2016 (openmopac.net)[38] and in-home utilities combine their capabilities and routinely exchange input and output structures of one another. The in-home software adopts the functions, procedures, and interfaces that are freely available in the ASE and SciPy.org computational chemistry libraries.[40,41] The visualization of molecular configurations, automatic generation of Z-matrices, visual investigation of the resulting geometries, and preparation of molecular graphics for the paper were carried out in VMD-1.9.1,[42] Gabedit-2.5,[43] and Avogadro-1.2.0 semantic chemical editor and analysis platform.[44]

**Results and Discussion**

Most of the chemical reactions have potential barriers that can trivially exceed a dozen of the k×T products. The barriers are associated with a necessity to first break the covalent bonds in the reactants before forming new chemical bonds in the products. The reaction barriers are a specific feature of nature that makes the surrounding world look and behave in its actual way. The presence of the potential barriers largely forbids brute-force simulating chemical transformations as we do with physical phenomena in which the changes of the gradients are much smoother. The high energy barrier means that the corresponding process is driven by the so-called rare events in the system, whereas its spontaneous modeling requires extensive, often computationally unaffordable sampling. To recapitulate, it is currently impossible to model chemical reactions without imposing a set of suppositions and restraints.

A chemical reaction path can be modeled assuming that a certain mechanism was previously hypothesized. In turn, it is possible to model a few anticipated mechanisms and compare their energetic profiles. By comparison of these profiles, the corresponding molecular geometries, and locations of the most essential saddle points, it is then possible to outline more and less probable paths of the reaction. The reaction path can be represented by a smoothly but rigidly moving internal coordinate, while other geometrical parameters are free to steadily accommodate the changes following the immediate wave function of the alternating system.

In the present work, we have the hypotheses that were inspired by the previously published experimental analyses.[23,27] While the $CO_2$ chemisorptions by the cation and the anion are likely rather independent chemical processes, the deprotonation of the α-carbon atom of the quaternary cation and grafting of the $CO_2$ molecule is sequential. No experimental insights are helping to decide which of them takes place first. The difference of thermodynamic potentials of the sub-reactions is also not a decisive factor because both sub-reactions take place in a single macroscopic observation. However, the potential energy barriers that can be extracted from the reaction path simulations can be highly insightful. The sub-reaction with a smaller barrier is naturally a more probable pathway of carboxylation. We simulated the mentioned processes in different orders and concluded that deprotonation (Figure 3) must occur before the carboxylation (Figure 4). If the $CO_2$ molecule approaches the tetrabutylphosphonium cation rather than tetrabutylphosphonium ylide it gives rise to no stable intermediate.

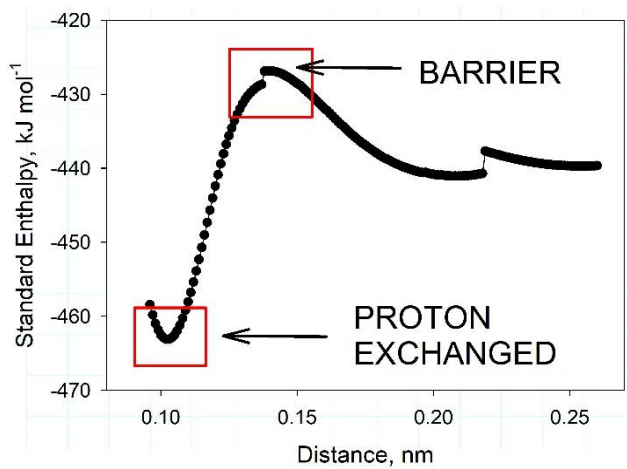

Figure 3. The recorded reaction path for the proton exchange reaction between the anion and the cation. The depicted reaction coordinate is a distance between the ring nitrogen atom of the 2-cyanopyrrolidine anion and the hydrogen atom linked to the α-carbon atom of the tetrabutylphosphonium cation. The initial reaction coordinate corresponds to the optimized geometry of the system with no restraints in the Z-matrix applied.

The barrier for carboxylation without prior deprotonation is fairly high amounting to 365 kJ mol⁻¹. Our analysis suggests that such stage is non-probable even upon significant heating, much over 333 K. In turn, the barrier after α-carbon deprotonation is as small as 13 kJ mol⁻¹ being in good agreement with the experimentally proven[25] sufficient heating up to 333 K to launch the chemical transformation. Furthermore, the shapes of the computed reaction path curves differ drastically.

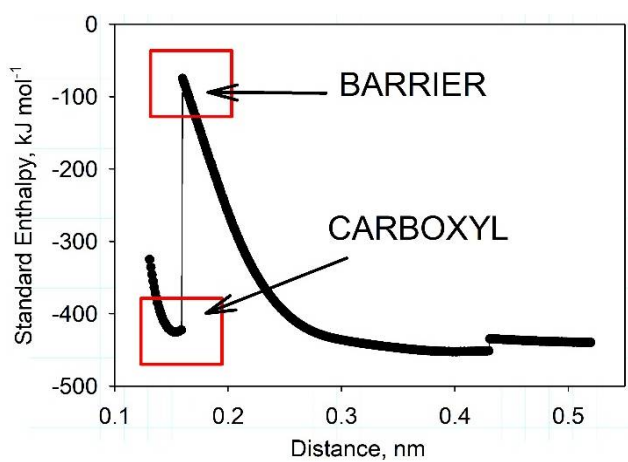

Figure 4. The reaction path for the carboxylation of the α-carbon atom of the tetrabutylphosphonium cation, assuming that no cation-anion proton transfer has taken place before. The depicted reaction coordinate is a distance between the carbon atom of $CO_2$ and one of the four α-carbon atoms of the tetrabutylphosphonium cation. The initial reaction coordinate corresponds to the optimized geometry of the system with no restraints in the Z-matrix applied.

Figure 5 shows $CO_2$ attachment to the deprotonated α-carbon atom and its eventual transformation into the negatively charged carboxyl group. This process occurs without any barrier, see the reaction coordinate of 0.238 nm. However, the approach of $CO_2$ to the methyl group of the TBP cation is associated with two distinct barriers, at 0.529 nm (hight of 14 kJ mol$^{-1}$) and 0.406 nm (height of 17 kJ mol$^{-1}$). Both of them are steric due to the collisions of $CO_2$ with the methylene groups of the TBA butyl chains. Since the density of the alkyl chains increases with the decrease of their distance to the center of TBP the barrier height increases proportionally. The overcoming of these steric barriers adds to the system's energy -7 and -10 kJ mol$^{-1}$ respectively. Recall that the reaction coordinate step was chosen to be 0.001 nm (much smaller than in other researches) to obtain the precise locations of the states that determine both covalent and non-covalent energetics of the chemisorption reactions.

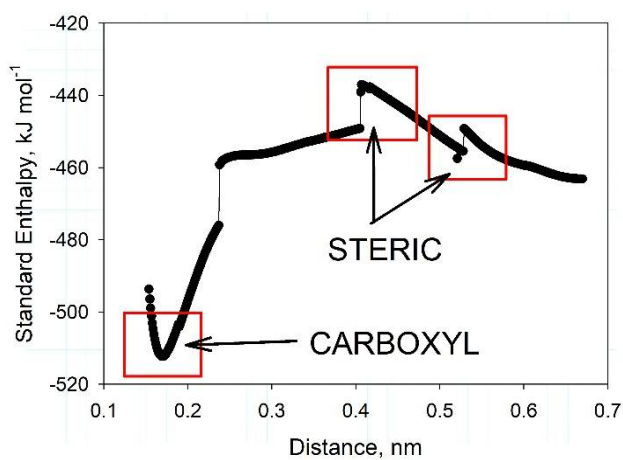

Figure 5. The reaction path for the carboxylation of the α-carbon atom of the tetrabutylphosphonium cation. This stage occurs after the proton exchange between the cation and the AHA. The reaction coordinate is a distance between the carbon atom of $CO_2$ and the α-carbon atom of the tetrabutylphosphonium cation.

The carbamate formation reaction proceeds even more smoothly (Figure 6) than the previously considered reaction at the α-methylene group at the cation and brings about $-10^2$ kJ mol$^{-1}$ to the enthalpy. The reaction on the 2-cyanopyrrolidine anion does not require activation energy unlike the reaction on the TBA cation.

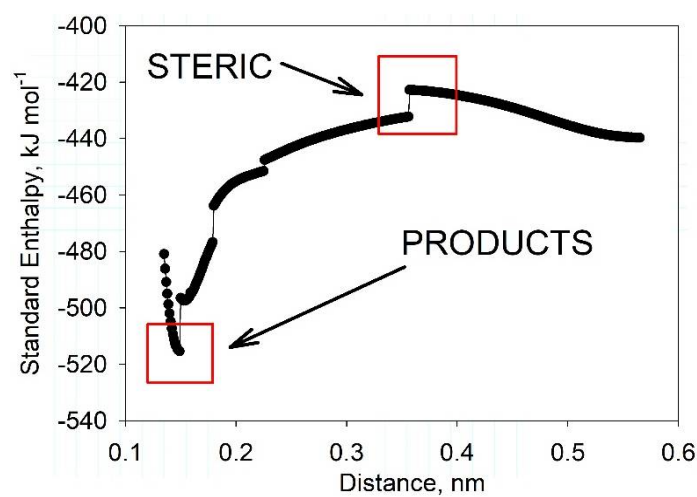

Figure 6. The reaction path for the carbamate formation reaction occurring at the electron-richest center of the 2-cyanopyrrolidine anion. The chosen reaction coordinate is a distance between the carbon atom of $CO_2$ and the ring nitrogen atom of the 2-cyanopyrrolidine anion.

Figure 7 visualizes the atomic configurations that correspond to the above-identified steric hindrances. Understanding these peculiar geometries is essential to assessing the feasibility of different quaternary cations for the $CO_2$ chemisorption. The long alkyl chains, such as those selected by Brennecke and coworkers,[23,25,27,37] are important to modulate the phase behavior of the resulting RTILs but they also increase the viscosity. High viscosity complicates the processing of RTILs in the experiment. The short chains decrease viscosity but increase the melting point of an RTIL. The present work suggests that the barriers associated with the flexibility of the alkyl groups are significantly high. In this context, the usage of asymmetric tails that help eliminate some steric barriers may be an interesting idea. At the reaction coordinate of 0.529 nm, the distance between the oxygen atom of $CO_2$ and the hydrogen atom of one of the methylene groups of TBP equals 0.25 nm (Figure 7). Furthermore, the simultaneous distance between the carbon atom of $CO_2$ and the nitrogen atom of AHA is 0.32 nm. These two collisions are responsible for the observed steric barrier. Similarly, the small distances between the $CO_2$ molecule and both ions of the RTIL can be observed at the reaction coordinate of 0.406 nm.

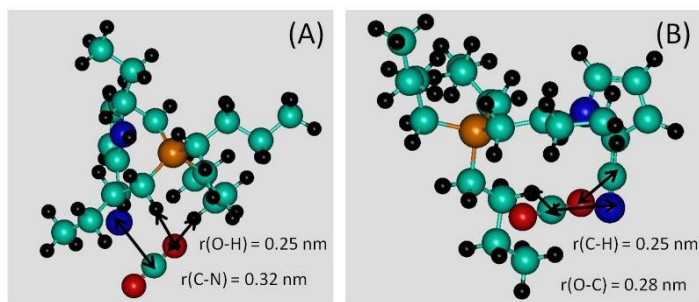

Figure 7. The atomic configurations along the reaction path corresponding to the sterical hindrances on the way of the $CO_2$ molecule to the α-carbon atom of the tetrabutylphosphonium cation. (A) The reaction coordinate of 0.529 nm and (B) the reaction coordinate of 0.406 nm. The oxygen atoms are red; the nitrogen atoms are blue; the carbon atoms are cyan; the hydrogen atoms are black.

Deprotonation of the TBA cation is principally different from the case of the TBP cation. This phenomenon has an evident chemical explanation since the TBA cation cannot form a corresponding ylide. Indeed, the formation of stable compounds with the nitrogen's valence of five is impossible. The electron transfer reaction is thermodynamically unfavorable, with the standard enthalpy change of $+15 \text{ kJ mol}^{-1}$ (Figure 8). In turn, the barrier of the reaction is $29 \text{ kJ mol}^{-1}$. The insignificant steric barrier of $2 \text{ kJ mol}^{-1}$ at the reaction coordinate of $0.22 \text{ nm}$ should be noticed. The overcoming of the mentioned steric barrier brings the system $5 \text{ kJ mol}^{-1}$. The above values are small enough to be exceeded by the thermal motion energy at room conditions.

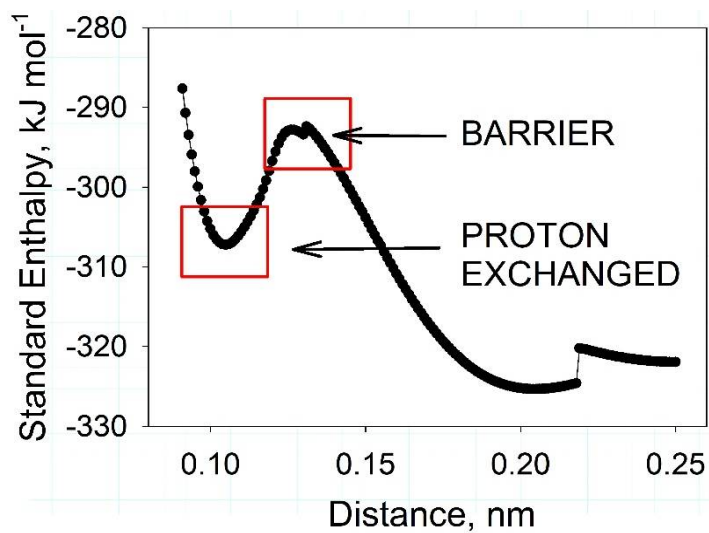

Figure 8. The reaction path for the proton exchange between the tetrabutylammonium cation and 2-cyanopyrrolidine anion. The chosen reaction coordinate is a distance between the ring nitrogen atom of the 2-cyanopyrrolidine anion and the hydrogen atom linked to the α-carbon atom of the tetrabutylammonium cation.

The electron transfer from the cation to the AHA is thermodynamically forbidden because the ammonium ylide is an unstable chemical entity. Nonetheless, it was desirable to characterize the entire reaction of the $CO_2$ capture. Figures 9-10 provide the corresponding reaction paths.

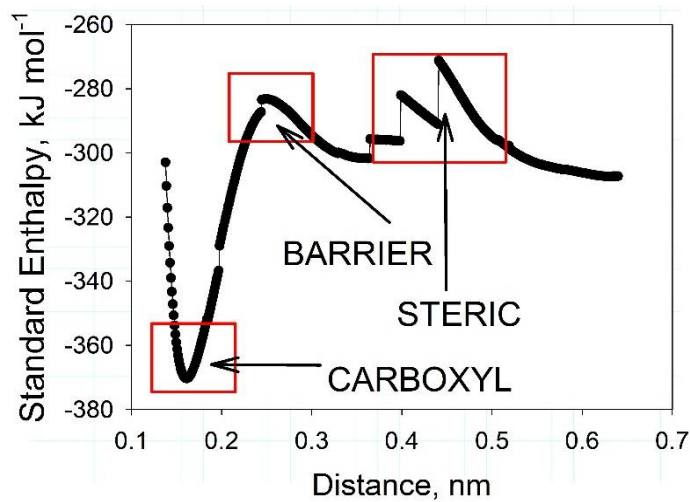

Figure 9. The reaction path for the carboxylation reaction taking place at the α-carbon atom of the tetrabutylammonium cation. The chosen reaction coordinate is a distance between the carbon atom of $CO_2$ and one of the α-carbon atoms of the tetrabutylammonium cation.

The reaction coordinate of 0.64 nm depicted in Figure 9 describes the state of the system in which the anion is protonated whereas the TBA cation is a free radical. As a result, the chemisorption barrier is rather small amounting to 19 kJ mol⁻¹. Compare this barrier to the case of the TBP cation in which the analogous barrier is 13 kJ mol⁻¹. Furthermore, the chemisorption reaction energy gain equals -68 kJ mol⁻¹, whereas the carboxylation of the TBP cation brings just -22 kJ mol⁻¹. To recapitulate, the carboxylation of the α-carbon atom of TBA is thermodynamically allowed but this reaction does not have a so convenient intermediate as phosphonium ylide. As a result, the potential usage of the TBA cation in the $CO_2$ capture is more cumbersome.

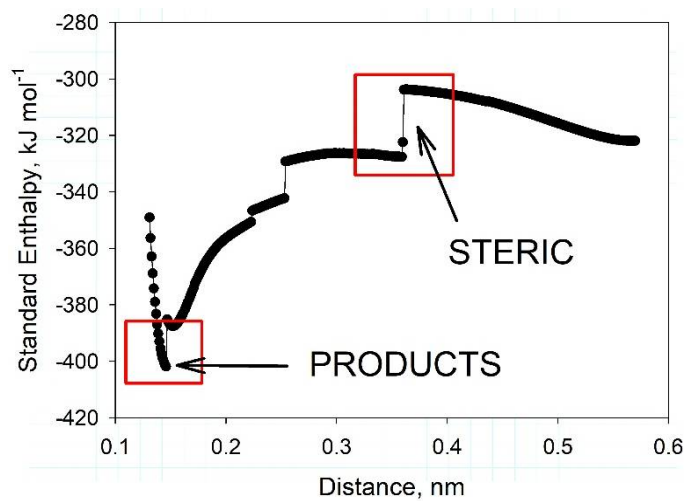

Figure 10. The reaction path for the carbamate formation reaction taking place at the electron-richest nitrogen atom of the 2-cyanopyrrolidine anion. The chosen reaction coordinate is a distance between the ring nitrogen atom of the 2-cyanopyrrolidine anion and the carbon atom of $CO_2$.

The carbamate formation reaction at the electron-rich N-site of the AHA in the case of the ammonium-based cation (Figure 10) is similar to the above-considered case of the TBA cation. The enthalpy gain is -74 kJ mol$^{-1}$. Note the steric barrier of 18 kJ mol$^{-1}$ observed at the reaction coordinate of 0.362 nm. It corresponds to the rearrangement of all components of the system (TBA, AHA, and $CO_2$) in response to the gradual movement of the $CO_2$ molecule towards the nitrogen atom of AHA. The overcoming of this barrier results in the system's enthalpy gain of -6 kJ mol$^{-1}$.

The identified in this work reaction barriers were linked to the specific transition states (Figure 11) through the unconstrained geometry optimization up to the local energy maximum. The initial geometries for this series of calculations were taken from the reaction path calculations by picking up a few highest-energy molecular conformations. Upon search along the reaction coordinate, the force constants were updated at every step. Each considered state exhibits a single imaginary vibrational frequency.

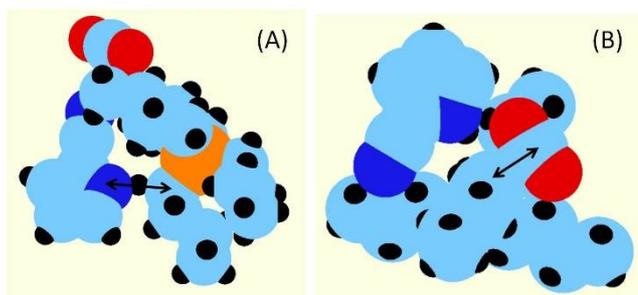

Figure 11. The transition state atomic configurations corresponding to the simulated chemical reactions. (A) Proton exchange between the tetrabutyphosphonium cation and the 2-cyanopyrrolidine anion. (B) Carboxylation of the deprotonated α-carbon atom of the tetrabutylammonium cation. The depicted black arrows display the key interatomic distances. The oxygen atoms are red; the nitrogen atoms are blue; the phosphorus atom is orange; the carbon atoms are cyan; the hydrogen atoms are black.

The deprotonation of the TBA reaction's transition state has an imaginary frequency of 605 cm$^{-1}$. The nitrogen(AHA)-hydrogen(TBA) distance in this transition state equals 0.131 nm. In turn, the carbon(TBA)-hydrogen(TBA) distance is 0.142 nm. The proton locates almost in the middle between nitrogen and carbon atoms in the described transition state. Importantly, upon proton exchange, the distance from the nitrogen atom to the deprotonating α-carbon atom of the TBA cation decreases down to 0.147 nm. This observation is in line with an emergence of the unpaired electron in the course of the proton exchange reaction. Compare the length of the nitrogen-carbon covalent bond to its equilibrium value in the unperturbed state, 0.153 nm.

The deprotonation of the TBA reaction's transition state vibrates at a similar imaginary frequency of 501 cm$^{-1}$. The nitrogen(AHA)-hydrogen(TBP) distance equals 0.139 nm, whereas the carbon-hydrogen distance equals 0.140 nm. In the transition state, the geometry of the TPB cation alters significantly. In particular, the phosphorus-$\alpha$-carbon atom covalent bond length decreases down to 0.172 nm. Compare this value to the length of this bond in the ground state of the TBP cation, 0.185 nm.

The transition state for the ammonium carboxylation exhibits imaginary vibrations at 113i cm$^{-1}$. In this configuration, the carbon($CO_2$)-carbon(TBA) distance equals 0.249 nm. The recorded distance agrees well with the van der Waals radii of the involved atoms. The saddle point locates at the distance where electron-electron repulsion should be expected. The O-C-O angle in the carbon dioxide molecule is 165 degrees. Compare this transition parameter with 131 degrees as computed at the same level of theory for the carboxylated product of the considered chemisorption reaction.

### Conclusions and Final Remarks

To recapitulate, the present work rationalizes carbon dioxide chemisorption by the quaternary ammonium and phosphonium cations coupled with the aprotic heterocyclic anion 2-cyanopyrrolidine. In each system, we considered three chemical processes that are involved in the $CO_2$ capture and identified their key features. It was found that the chemisorption reaction is initiated by the donation of the proton from the cation to the anion. This process is thermodynamically favorable in the case of the phosphonium cation and thermodynamically unfavorable in the case of the ammonium-based cation. The chemisorption event is expected to fail if the $CO_2$ molecule approaches the $\alpha$-carbon atom of the cation before its deprotonation.

This conclusion was drawn from the analysis of the reaction barriers upon the simulation of the competitive processes.

The sterical hindrances introduced by the saturated alkyl chains of the quaternary ammonium- and phosphonium-based cations play a significant role in the $CO_2$ chemisorption. The role and the scale of the hindrances are revealed in the reaction path diagrams. The hindrances are more essential upon carboxylation than upon deprotonation. In turn, the $CO_2$ capture according to the carbamate formation mechanism is also influenced by the alkyl chains of the cations. This feature might have been expected since the cation-anion attraction is by default stronger than cation-$CO_2$ and anion-$CO_2$ non-bonded (dispersion and electrostatic) attractions.

The chemisorption of $CO_2$ by the tetrabutylphosphonium cation involves the formation of the tetrabutylphosphonium ylide. Subsequently, the $CO_2$ molecule gets grafted. Thanks to the double carbon-carbon bond, carboxylation takes place with essentially no barrier. In turn, the tetrabutylammonium ylide is a metastable sub-product. It was possible to detect an energy minimum corresponding to it, but such a compound is highly reactive due to the inability of nitrogen to afford the valence of five. As a result of its chemical activity, the ylide reacts with the $CO_2$ molecule readily and forms the carboxyl group at the α-carbon atom of the ammonium-based cation. The 2-cyanopyrrolidine participates in the carbamate formation reaction which is associated with a very small potential barrier. The carbamate production at the anion's reaction site is almost independent of the counterion of the RTILs.

The comparison of the different simulated stages of the $CO_2$ chemisorption reveals the following facts. The $CO_2$ capture by the phosphonium-based cation is more thermodynamically favorable than by the ammonium-based cation. Furthermore, the deprotonation of the phosphonium-based cation is more favorable than the deprotonation of the ammonium-based cation. The carbamate formation reaction is more favorable than both of the above carboxylation reactions. Our findings are in qualitative agreement with the available experimental results.[25] In particular, the experiments suggest that only the heterocyclic anion captures $CO_2$ at standard conditions, whereas the phosphonium-based cation participates in its respective reaction at above 333 K. Indeed, the observed different behavior of the ions is fingerprinted in the calculated reaction barriers. The ammonium-based cation can also be probed for the $CO_2$ capture, although somewhat deteriorated performance is to be expected in comparison with the currently experimentally probed phosphonium-based cations.[21]

The reported computational results supplement the available experimentally derived chemical insights by providing a higher-resolution vision of the involved chemisorption reactions. The obtained in-silico data support our understanding of the α-carbon atoms of the quaternary ammonium- and phosphonium-based cations as the prospective $CO_2$ reaction sites. The reported analysis fosters a fundamental understanding of the $CO_2$ chemisorption reactions and stipulates chemical efforts to improve the existing $CO_2$ scavengers.

**Acknowledgments**

All reported numerical simulations have been conducted at the P.E.S. computational facility.

**Conflict of interest**

The author hereby declares no existing financial interests concerning these research studies.

## Authors for correspondence

All correspondence regarding the content of this paper shall be directed through electronic mail to vvchaban@gmail.com.